# Discovery of Multiple, Ionization-Created Anions in Gas Mixtures Containing $CS_2$ and $O_2$

Daniel P. Snowden-Ifft

*Department of Physics, Occidental College, Los Angeles CA  90041*

The use of negative ions in TPCs has several advantages for high-resolution rare-event detection experiments.  The DRIFT experiment, for example, has taken full advantage of this technique over the past decade in a directional search for dark matter.  This paper focuses on the surprising discovery of multiple species of ionization-created $CS_2$ anions, called minority carriers, in gas mixtures containing electronegative $CS_2$ and $O_2$, identified by their slightly different drift velocities.  Measurements of minority carriers in gas mixtures of $CS_2$, $CF_4$ and $O_2$ are reported in an effort to understand the nature of these charge carriers.  Regardless of the micro-physics however, this discovery offers significant practical advantages for experiments such as DRIFT, where the difference in arrival time may be used to fiducialize the original ionization event without an external start pulse.

Time Projection Chambers (TPCs) have been used extensively for decades [1] in a wide variety of particle detection experiments.  In 2000 Martoff [2] suggested the use of $CS_2$ anions in TPCs as a way of overcoming the high temperature diffusion of electrons between ionization and detection without the use of a magnetic field.  The "magic" of $CS_2$ is room temperature diffusion [3, 4] in all 3 dimensions combined the ability of high electric fields in the detection region to strip the electrons off of the $CS_2$ anions allowing for normal electron avalanche.  The Directional Recoil Identification From Tracks (DRIFT) experiment has taken full advantage of this technique [5].  Additional advantages of negative ion TPCs are stability over long time periods, low cost electronics due to slow drift speeds and a high tolerance for gas impurities.  The focus of this paper is on two additional significant features of $CS_2$ anion drift, the discovery of multiple species of drifting anions and their enhancement via the addition of small amounts of $O_2$.

The apparatus used in this experiment was fully described in [4].  Briefly, UV photons from a flashlamp were focused to a small spot on an Al cathode where photoelectrons were generated in a 40 Torr $CS_2$ gas mixture.  The resulting $CS_2$ anions were then drifted 15.24 cm in a uniform drift field to a Multi-Wire Proportional Chamber (MWPC) where amplification and readout occurred on 20 µm stainless steel anode wires.  As with other previous and similar experiments [2, 3, 6-8] a single peak, characterized by a drift time and width, was observed.  Mobility and diffusion results were reported for 40 Torr $CS_2$ and 30-10 Torr $CS_2$-$CF_4$ gas mixtures.  The mobilities of the $CS_2$ anions, $\mu$ and $\mu_0$ (the mobility at a reference temperature of 0C), are defined in the following equation,

$$v_{drift} \equiv \mu\left(\frac{E}{p}\right) \equiv \mu_0 \sqrt{\frac{T}{T_0}} \left(\frac{E}{p}\right) \tag{1}$$

where $E$ is the drift field, $p$ is the pressure, $T$ is the temperature of the gas and $T_0 = 273.15$ K.  As expected the $CS_2$ anions thermally diffused at the room temperature of the gas. For purposes of nomenclature this single peak will be called the I peak, for ionization.

Not discussed in [4] was the observation of 2 small peaks arriving before the I peak. These peaks sporadically appeared ~2.5% ahead, traveling faster and labeled S, and ~5% ahead, labeled P, of the I peak, had a similar width to the I peak but were significantly smaller in amplitude, ~0.1%, in comparison to the I peak. Only through averaging, afforded by the repeatability of this apparatus, were these peaks visible. Because of their small size these peaks were labeled minority peaks and the (unknown) charge carriers responsible for them, minority carriers. The observations strongly suggested that two different species of electron carriers were being created at the site of the photoelectron generation. These then drifted simultaneously but with slightly different, but stable, mobilities over 5-20 ms timescales, finally avalanched at the MWPC. Multiple drifting $CS_2$ "isotopes" from differing anion formation rates from Rydberg atom electron transfer have previously been observed [9] but this is the first observation of such an effect from an ionization process.

The similarity of the mobilities, and the fact that they avalanche suggest that they are variants of the normal $CS_2$ anions, the I carriers. There is significant literature available on the formation of $CS_2$ anions, see [9] for a summary. In its neutral ground state $CS_2$ is a linear molecule; after an electron is attached, though, it bends. This state is unstable on time scales of molecular vibrations but it can be made stable by the Block-Bradbury attachment mechanism involving a buffer gas. Presumably $CS_2$ acts as its own buffer gas in the case of pure $CS_2$ anion drift. Molecular "isotopes" of $CS_2$ with different attachment rates [9] are a possible explanation for the minority carriers. However all of the "isotopes" of $CS_2$ are heavier so, since the minority carriers travel faster than the I carriers, that explanation would involve a finely-tuned and compensating smaller drift cross section and or capture rates. In the low field limit [10],

$$\mu_0 \propto \sqrt{\frac{1}{m} + \frac{1}{M}}\frac{1}{\sigma} \qquad (2)$$

where $m$ is the mass of the drifting anion, $M$ is the mass of the gas particles and $\sigma$ is the collisional cross section. An alternative explanation may be that the minority carriers are atomic mass 76 $CS_2$ anions bent at different angles and therefore having different collisional cross sections and mobilities. More and better information is needed to solve this puzzle.

The main practical benefit of the minority peaks is, analogous to the fiducialization of earthquakes locations using S and P waves, trigger-less fiducialization of events in the gas. DRIFT, for instance, is heavily limited in its search for dark matter recoils by radon progeny recoils from its central cathode [DIIdLimit]. Without a trigger DRIFT is forced to use the RMS in time of the observed events as a crude method of fiducialization but this entails a factor of ~25 reduction in its detection efficiency. Were it not for their small size the ability to fiducialize events with the minority peaks would have a huge impact on DRIFT and other rare-event TPCs.

Fortunately, the addition of a small amount of $O_2$ to the gas mixture has been found to greatly increase the size of the minority peaks, as presented in Fig. 1.

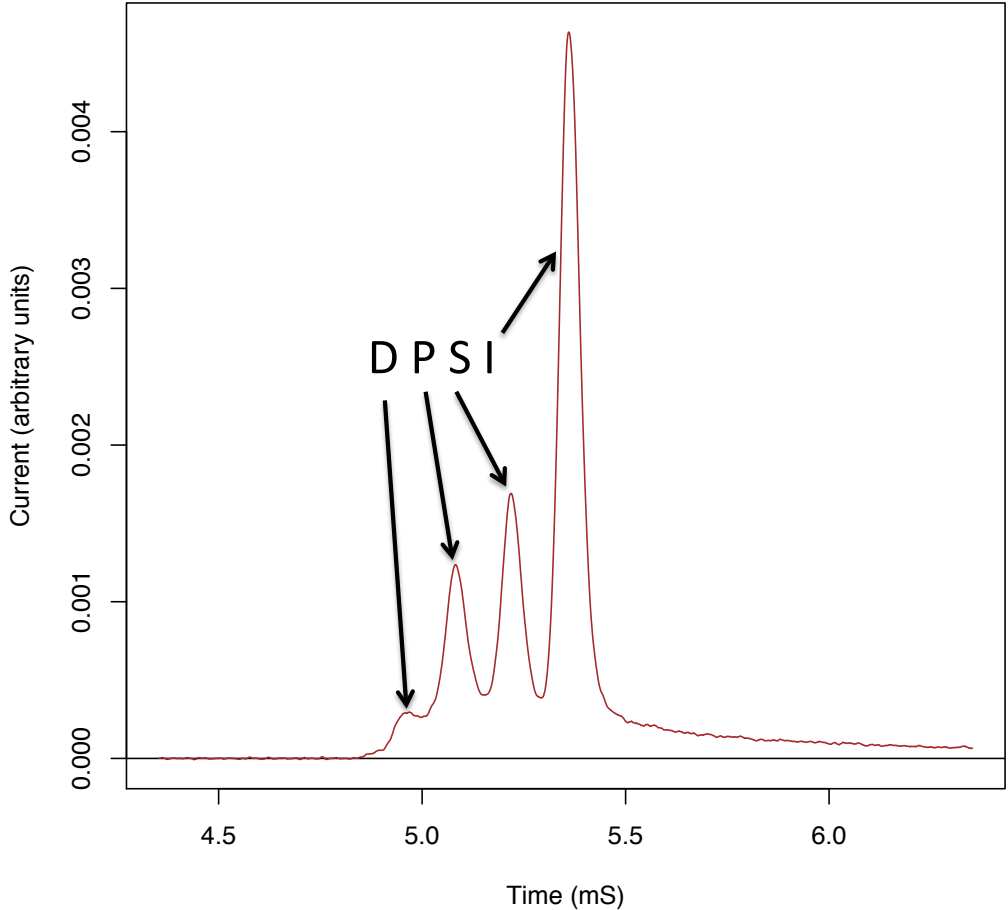

FIG. 1 – The arrival time distribution of negative ions after a 15.24 cm drift in a 273 V/cm drift field in a mixture of 30 Torr $CS_2$, 10 Torr $CF_4$ and 1 Torr $O_2$. Followng earthquake fidudialization and spectroscopic notation the minority peaks are labeled as shown.

It will be shown below that the I carriers in mixtures with small additions of $O_2$ are the same as the I carriers observed in previous experiments. Only 3 minority peaks have been identified thus far. The appearance of the D peak is ephemeral so this paper will focus solely on the properties of the S and P peaks.

It is now thought that the appearance of the miniscule S and P minority peaks in [4] was due to a small concentration of $O_2$ in the gas due to outgassing. Previous mobility measurements have been performed with gas mixtures of $CS_2$ and $CF_4$, Ar, Xe, $CH_4$, He, Ne and $CO_2$ [2-4, 6, 8] and there have been no reports of significant minority peak formation in those gas mixtures. That they appear with the addition of $O_2$ is a clue, a convenient means of studying them and a practical means for taking advantage of them. The purpose of this paper is to characterize the behavior of the S and P minority carriers to shed light on their identity and formation mechanism.

Since the widths of the peaks are identical the height serves as a convenient proxy for the total amount of post-avalanche charge in the peak. The first experiment for this paper to study the charge distribution of the minority peaks used a UV spot positioned on the cathode over one of the wires at a fixed drift field but with varying concentrations of $O_2$. Data were taken first with a base gas of 40 Torr $CS_2$. $O_2$ was then added to the gas mix in ~0.2 Torr increments up to ~1.0 Torr added $O_2$. The experiment was then repeated with a 30-10 Torr $CS_2$-$CF_4$ gas mixture as the base. The increase in gas pressure and changing gas mixture changed the gas gain on the wire. For this reason the ratio of the heights of the minority peaks relative to the I peak is displayed in Fig. 2.

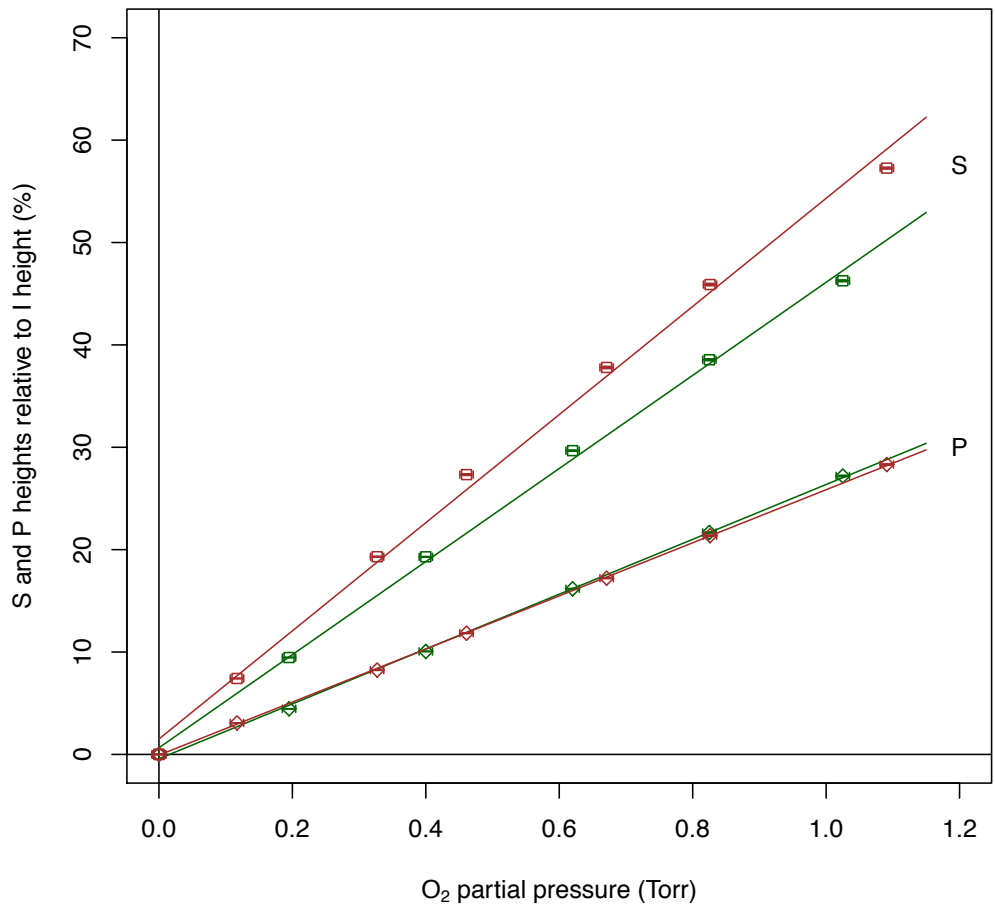

FIG. 2 – The size of the S and P minority peaks relative to the size of the I peak for varying partial pressures of $O_2$ added to either 40 Torr $CS_2$, shown in brown, or 30-10 Torr $CS_2 - CF_4$, shown in green. Both measurements were done with a 15.24 cm drift distance. The $CS_2$ measurements were done in a drift field of 299 V/cm while the $CS_2$-$CF_4$ measurements were done in a field of 314 V/cm. A linear fit is shown for each data set.

As Fig. 2 makes clear the formation of the S and P minority carriers clearly requires the presence of $O_2$. The linear dependence shown in Fig. 2 suggests a formation mechanism involving only a single $O_2$ molecule, consistent with the Block-Bradbury attachment mechanism. In the case of pure $CS_2$ a $CS_2$ molecule would act as the "buffer" gas for the electron attachment to another $CS_2$ molecule. However, the percentages shown in Fig. 2 for the P peak are identical for the different gas mixtures, with different $CS_2$ partial pressures, but different for the S peaks suggesting that the formation mechanisms for the S and P carriers are different.

The S carriers appear to have a finite lifetime. In this experiment a UV spot was, again, positioned on the cathode over one of the wires but at fixed concentrations of $O_2$ and with varying drift fields and therefore varying drift times. Diffusion affects the height of the peaks which is, in turn, heavily affected by the drift field so, again, the S and P peak heights were compared to I peak height. No change was found in the P peak height. The S peak height, however, was found to systematically decrease with increasing drift time. Fig. 3 shows a semi-log plot of the ratio of S peak height to I peak height as a function of drift time. The linear dependence shown on this semi-log plot suggests that S carriers have a finite lifetime. An alternative hypothesis, that the decrease is inversely related to drift fields, has been ruled out from DRIFT data taken with $O_2$ present and will be discussed in a separate paper.

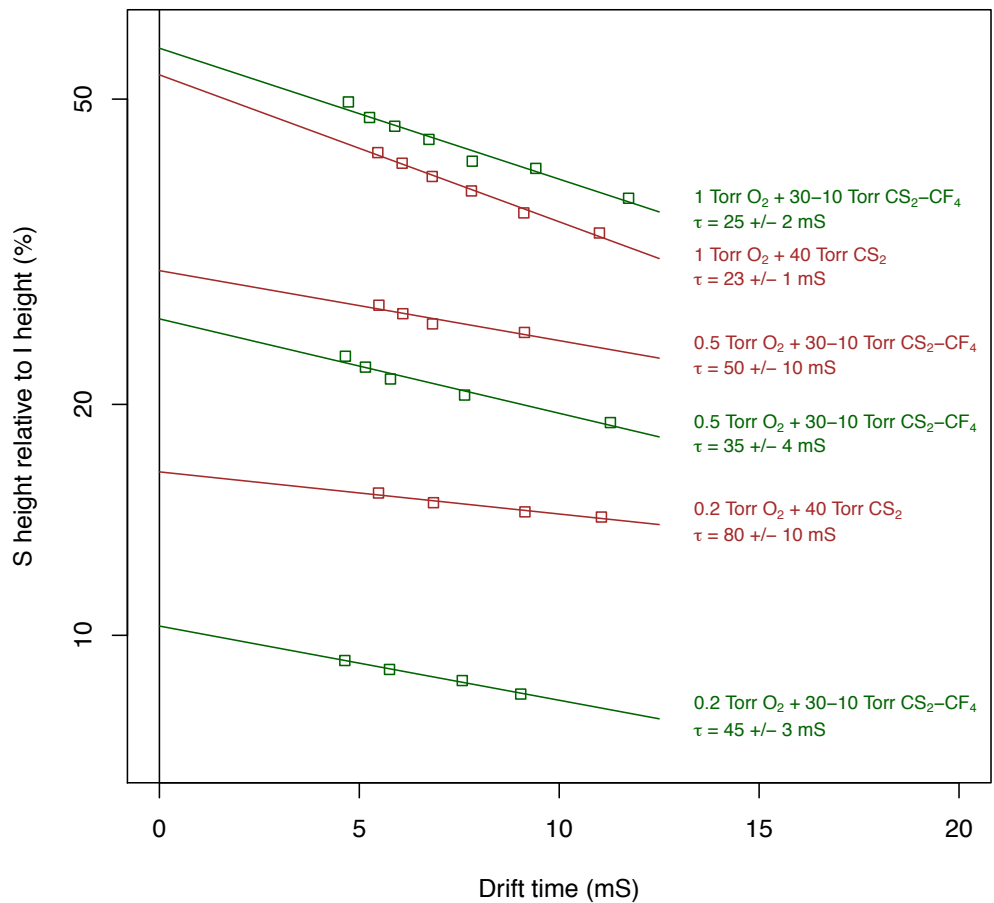

FIG. 3 – The size of the S peak relative to the size of the I peak for varying drift times in various gas mixtures.

The drift times of the P, S, and I peaks as a function of $O_2$ partial pressure and gas mixture were measured using the data set described above. Using the procedures described in [4] the mobility of each peak for each different $O_2$ partial pressure was calculated. The calculated measured mobilities were normalized to mobilities at 0 C, $\mu_0$, using Eq. (1). The results are shown in Fig. 4.

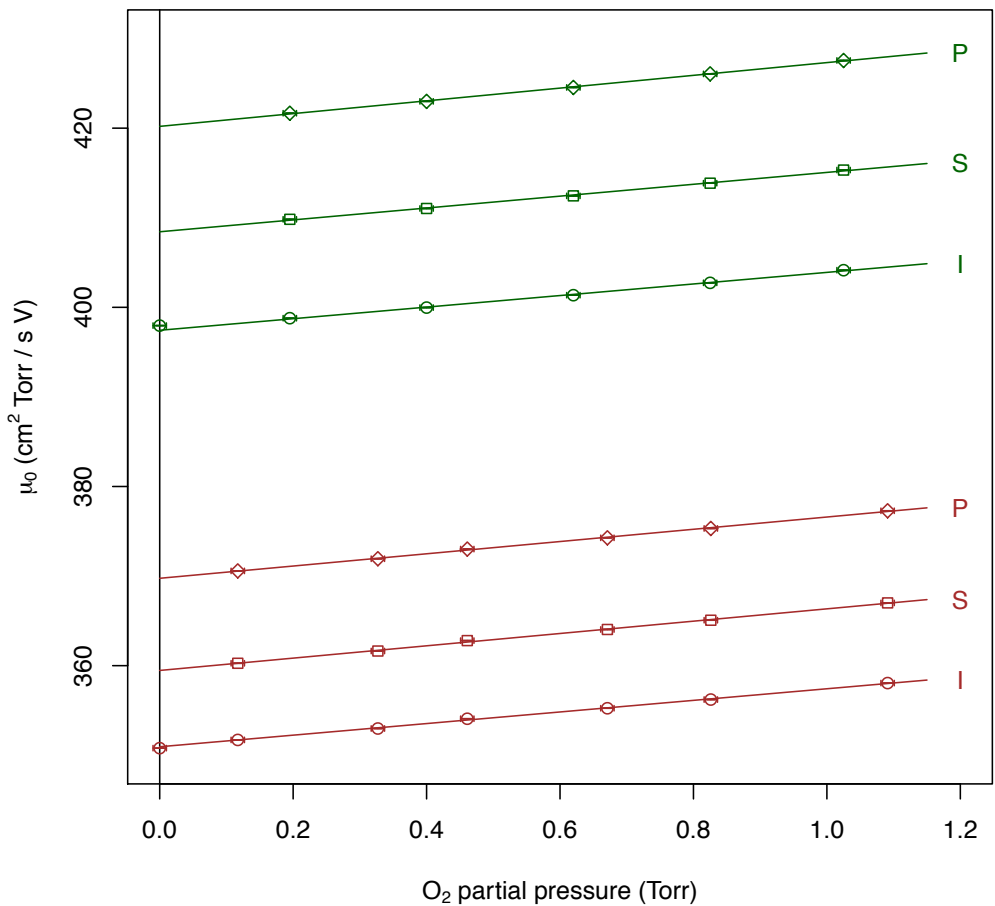

FIG. 4 – Mobilities, corrected to 0 C, for the I and S and P minority carriers in a base gas of 40 Torr $CS_2$, shown in brown, and 30-10 Torr $CS_2 - CF_4$, shown in green.

The results at 0.00 Torr $O_2$ agree well with the results in [4], particularly in light of the systematic effects discussed therein. The mobilities increase with increasing $O_2$ partial pressures since the average mass of a gas molecule and interaction cross section decrease with increasing $O_2$ content, see Eq. (2). These trends were fit with simple linear regressions and are summarized in Table 1. Importantly the fits to the I peak data *do not* include the points at 0.00 Torr $O_2$. As can be seen, though, the data at 0.00 Torr $O_2$ agree well with a simple linear extrapolation of the data at higher partial pressures of $O_2$. This, in turn, suggests that the I peak in minority peak data is the same peak as measured in [4] and elsewhere. That in turn suggests that the formation of the I peak, in mixtures or not, does not require $O_2$.

Table 1 – Summary of P, S and I mobility parameters for linear fits to $O_2$ partial pressures. All mobility parameters are quoted at 0 C.

|  | P Intercept ($cm^2$ Torr/s V) Slope ($cm^2$ Torr/s V/Torr $O_2$) | S Intercept ($cm^2$ Torr/s V) Slope ($cm^2$ Torr/s V/Torr $O_2$) | I Intercept ($cm^2$ Torr/s V) Slope ($cm^2$ Torr/s V/Torr $O_2$) |
|---|---|---|---|
| 30-10 Torr $CS_2$-$CF_4$ | 420.20 +/- 0.08 | 408.4 +/- 0.1 | 397.4 +/- 0.1 |
|  | 7.1 +/- 0.1 | 6.6 +/- 0.2 | 6.5 +/- 0.1 |
| 40 Torr $CS_2$ | 369.75 +/- 0.08 | 359.46 +/- 0.08 | 350.95 +/- 0.08 |
|  | 6.8 +/- 0.1 | 6.9 +/- 0.1 | 6.5 +/- 0.1 |

As a test of minority peak formation by other sources of ionization and a demonstration of minority peak fiducialization a $^{210}$Po alpha source was introduced in the vacuum vessel near the detector. Alpha particles traveling at ~45° from the normal to the MWPC (the z coordinate) and roughly perpendicular to the anode wire direction (the x coordinate) were selected. As described in [4] the wires of the detector are grouped together to form 8 output lines. For this experiment 4 output lines, with wires spaced at 4 mm, were monitored. An example event is shown in Fig. 5. As the alpha particle passed over a wire, a segment of ionization with $\Delta x \sim \Delta z \sim$ 2mm was created. The electrons were then captured and drifted down to the wire. The repeating structure evident in Fig. 5 is due to the grouping of the wires. Minority peaks are clearly visible in the later peaks. A closer look reveals that the minority peaks *are* visible for earlier peaks but their separation from the I peak decreases with distance to the MWPC, as expected,

$$\Delta t_{S,P} = \left(\frac{1}{\mu_I} - \frac{1}{\mu_{S,P}}\right)\left(\frac{P}{E}\right)z \qquad (3)$$

where z is the distance between the ionization and the MPWC. For the earliest ionization, the minority peaks are completely hidden within the I peak. The large spike in ionization at $t = 0$ is caused by the alpha particle traveling through the MWPC and the rapid accumulation of charge in the high-field, 1.1 cm, grid-anode gap. This distinct, rapid collection of charge fiducializes the alpha particle; the ionization at $t = 0$ was created at $z = 0$. The arrival time of an I peak relative to this $t_0$ provides an accurate measure of the average z of the ionization segment, $z_0$. On the other hand the separation between the S and P peaks and the I peak provides another, trigger-less, estimation of the average z of the ionization segment, $z_{MP}$ through Eq. (3). Fig. 6 shows the relation between $z_0$ and $z_{MP}$.

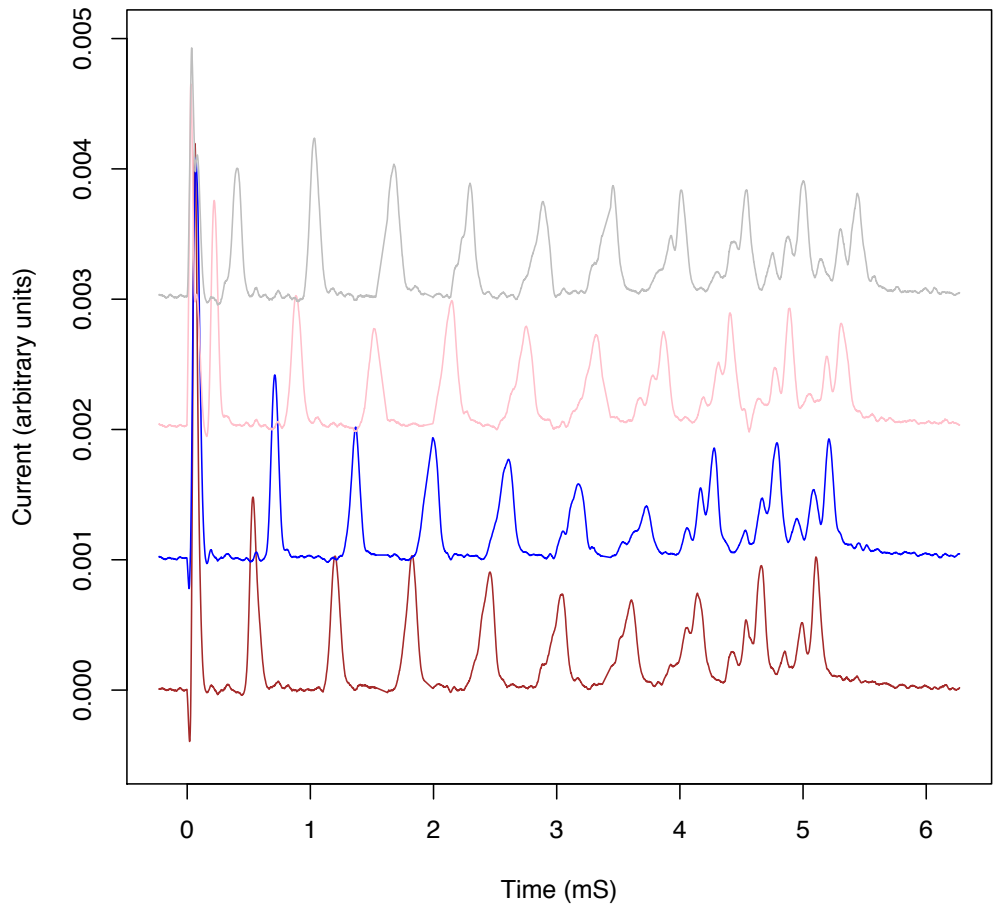

FIG. 5 – A $^{210}$Po alpha particle event in the detector filled with 40 Torr $CS_2$ and 0.94 +/- 0.01 Torr of $O_2$ in drift field of 293 V/cm at T = 23.4 C. As discussed in the text the rapid collection of charge at the beginning of the event provides a $t_0$ to fiducialize the alpha particle. The inferred currents on the 4 monitored lines (spaced at 4 mm) were offset by 0.001 for presentation purposes.

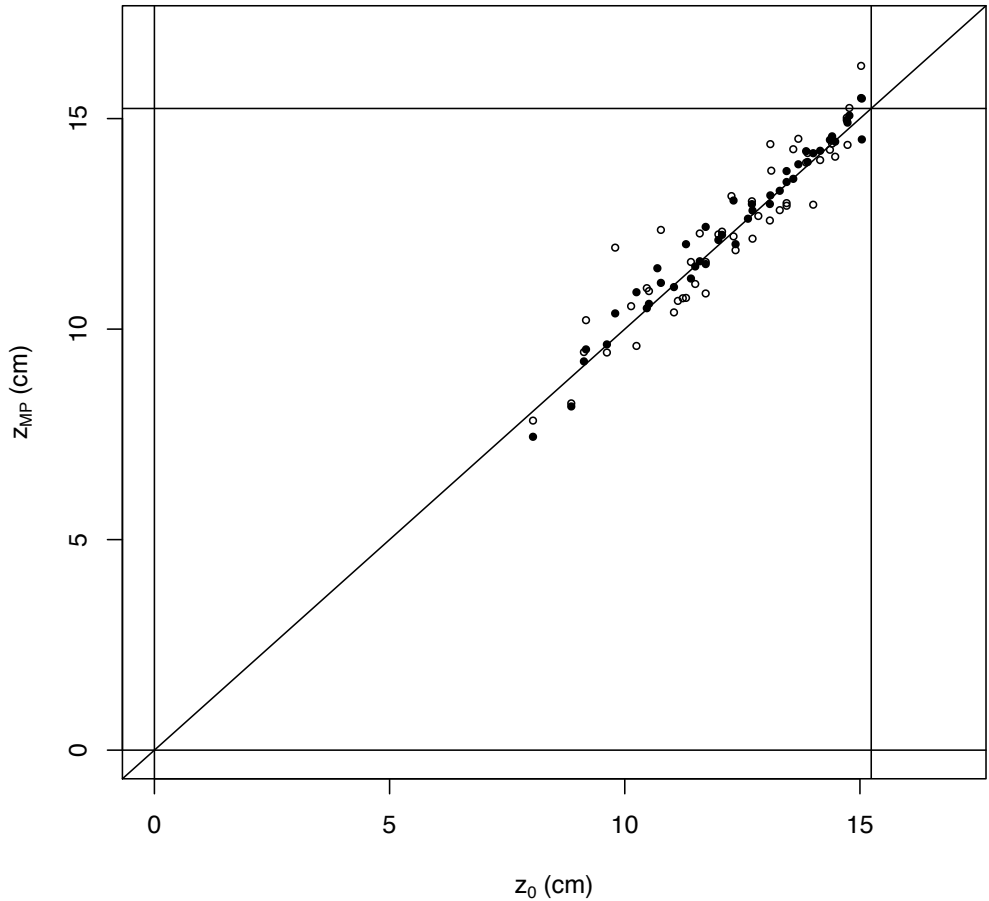

Fig. 6 – The distance, $z$, between an alpha created ionization segment and the MWPC measured in 2 ways. In the first the time difference between the arrival of the I peak and a $t_0$, discussed in the text, provides one measure of $z$, $z_0$. In the second the time difference between either the S peak or the P peak and the I peak, Eq. (3) and the values in Table 1 provide a second measure of $z$, $z_{MP}$. $z_0$ vs $z_{MP}$ is shown on the plot with the S peak data shown with open circles and the P peak data shown with filled circles. The vertical and horizontal lines at 15.24 cm show the $z$ dimension of the detector. At least one of the measured alphas hit the cathode while over one of the measurement lines. The line drawn with a slope of 1 shows the strong agreement between these two ways of measuring $z$. The measurement of $z_{MP}$ suffers from lower resolution. Assuming the resolution of $z_0$ is much greater than $z_{MP}$ the inferred *RMS* for the measurement of $z$ from the S peak is $\sigma_S = 0.67$ cm and for the P peak $\sigma_P = 0.33$ cm, the difference due simply to the larger separation for the P peak. It also suffers from the inability to measure events close to the MWPC. However, the measurement of $z$ using the minority peaks has the distinct advantage of not requiring a trigger.

For this analysis the position of all the peaks were measured by eye so for ionization segments near the detector no measurement was possible. A sophisticated peak-finding algorithm would certainly improve this result. For this result the ionization segments agreement between the two measurements is quite good $z >\sim 7.5$ cm. The importance of this result is that for a single ionization segment created, for instance by a neutron or WIMP recoil in DRIFT, where no trigger is possible, a measurement of $z_{MP}$ is still possible. For DRIFT this will allow the removal of its only known backgrounds, events from the MWPC and central cathode, without undue loss of effective volume.

In conclusion, the addition of $O_2$ to gas mixtures including $CS_2$ and in the presence of ionization produces at least 2 distinct species of drifting anions in addition to the normal, drifting $CS_2$ anion. The various measurements presented above on these minority carriers will hopefully spur further research into their identity. Aside from their inherent interest, the existence of minority carriers is already proving to have enormous practical importance for experiments such as DRIFT. Other high-resolution rare-event TPC experiments could also benefit from this technique.

This work is supported by the National Science Foundation. The author is very grateful to the DRIFT collaboration and, in particular, Alex Murphy for useful discussions and comments in the development of this work.